\documentclass[twocolumn,aps,prl,showpacs,superscriptaddress,floatfix]{revtex4}

\usepackage[dvips]{color} 
\usepackage{graphicx} % Include figure files
\usepackage{bm}% bold math

\begin{document}

%\preprint{}

\title{Semi-spheroidal Quantum Harmonic Oscillator}

\author{D. N. Poenaru}
\email[]{poenaru@fias.uni-frankfurt.de}
%\homepage[]{http://www.th.physik.uni-frankfurt.de/{\verb+~+{poenaru}}

\affiliation{Frankfurt Institute for Advanced Studies,
J. W. Goethe Universit\"at, 
        Max-von-Laue-Str. 1,    D-60438 Frankfurt am Main,   Germany}
\affiliation{
Horia Hulubei National Institute of Physics and Nuclear
Engineering (IFIN-HH), \\P.O. Box MG-6, RO-077125 Bucharest-Magurele, Romania}

\author{R. A. Gherghescu}

\affiliation{Frankfurt Institute for Advanced Studies,
J. W. Goethe Universit\"at, 
        Max-von-Laue-Str. 1,    D-60438 Frankfurt am Main,   Germany}
\affiliation{
Horia Hulubei National Institute of Physics and Nuclear
Engineering (IFIN-HH), \\P.O. Box MG-6, RO-077125 Bucharest-Magurele, Romania}

\author{A. V. Solov'yov}
\affiliation{Frankfurt Institute for Advanced Studies,
J. W. Goethe Universit\"at, 
        Max-von-Laue-Str. 1,    D-60438 Frankfurt am Main,   Germany}

\author{W. Greiner}
\affiliation{Frankfurt Institute for Advanced Studies,
J. W. Goethe Universit\"at, 
        Max-von-Laue-Str. 1,    D-60438 Frankfurt am Main,   Germany}

\date{\today}

\begin{abstract}
A new single-particle shell model is derived by solving the Schr\"odinger
equation for a semi-spheroidal potential well. Only the negative parity
states of the $Z(z)$ component of the wave function are allowed, so that
new magic numbers are obtained for oblate semi-spheroids,
semi-sphere and prolate semi-spheroids. The semi-spherical magic numbers are
identical with those obtained at the oblate spheroidal superdeformed shape:
2, 6, 14, 26, 44, 68, 100, 140, ... The superdeformed prolate magic numbers
of the semi-spheroidal shape are identical with those obtained at the
spherical shape of the spheroidal harmonic oscillator: 2, 8, 20, 40, 70,
112, 168 ...
\end{abstract}


\maketitle

The spheroidal harmonic oscillator have been used in various branches of
Physics. Of particular interest was the famous single-particle Nilsson
model \cite{nil55dv} very successful in Nuclear Physics and its variants 
\cite{kni84prl,cle85prb,rei93zpd} for atomic clusters.
Major spherical-shells $N=2, 8, 20, 40, 58, 92$ have been found
\cite{kni84prl} in the mass spectra of sodium clusters of $N$ atoms per
cluster, and the Clemenger's shell model \cite{cle85prb} 
was able to explain this sequence of spherical magic numbers.

In the present paper we would like to write explicitly the analytical
relationships for the energy levels of the spheroidal harmonic oscillator
and to derive the corresponding solutions for a semi-spheroidal harmonic
oscillator which may be useful to study atomic cluster deposited on planar
surfaces.

For spheroidal equipotential surfaces, generated by a potential with
cylindrical symmetry the states of the valence electrons were
found \cite{cle85prb} 
by using an effective single-particle Hamiltonian with a potential
\begin{equation}
V=\frac{M\omega^2_0 R^2_0}{2}\left[\rho^2\left(\frac{2
+\delta}{2-\delta}\right)^{2/3}
+z^2\left(\frac{2-\delta}{2+\delta}\right)^{4/3}\right]
\end{equation}
In order to get analytical solutions we shall neglect an additional term
proportional to $({\bf l}^2 - \langle {\bf l}^2 \rangle_n)$. We
plan to include in the future such a term which needs a numerical solution.
K. L. Clemenger introduced the deformation $\delta$
by expressing the dimensionless two semiaxes (in units of the radius 
of a sphere with the same volume, $R_0=r_sN^{1/3}$, where $r_s$ is the 
Wigner-Seitz radius, 2.117~{\AA} for Na \cite{bra89prb,yan95prb}) as
\begin{equation}
a=\left(\frac{2-\delta}{2+\delta}\right)^{1/3} \; \; ; \; \;
c=\left(\frac{2+\delta}{2-\delta}\right)^{2/3} 
\label{clem}
\end{equation}
The spheroid surface equation
in dimensionless cylindrical coordinates $\rho$ and $z$ is given by
\begin{equation}
\frac{\rho^2}{a^2}+\frac{z^2}{c^2}=1
\end{equation}
where $a$ is the minor (major) semiaxis for prolate (oblate)
spheroid and $c$ is the major (minor) semiaxis for prolate (oblate)
spheroid. Volume conservation leads to $a^2c=1$.

One can separate the variables in the Schr\"odinger equation, $H\Psi =
E\Psi$, written in cylindrical coordinates. As a result the wave function 
\cite{ras58pr,vau73pr} may be written as
\begin{equation}
\Psi (\eta, \xi, \varphi) = \psi^m_{n_r} (\eta) \Phi_m(\varphi) Z_{n_z}
(\xi)
\end{equation}
where each component of the wave function is ortonormalized 
leading to
\begin{equation}
\Phi_m (\varphi) = e^{im\varphi}/\sqrt{2\pi}
\end{equation}
\begin{eqnarray}
\psi (\eta) & = N^m_{n_r} \eta ^{|m|/2} e^{-\eta /2} L_{n_r}^{|m|} (\eta)
\nonumber \\ & 
N^m_{n_r}=\left(\frac{2n_r !}{\alpha_\perp (n_r + |m|)!}\right)^{1/2}
\end{eqnarray}
in which $\eta = R_0^2 \rho ^2 /\alpha ^2 _{\perp}$ and
the quantum numbers $m=(n_{\perp} -2i)$ with $i=0, 1, ...$ up to 
$(n_{\perp} -1)/2$ for an odd $n_{\perp}$ or to $(n_{\perp} -2)/2$
for an even $n_{\perp}$. $L_n^m(x)$ is the associated Laguerre
polynomial and the constant $\alpha _{\perp} =\sqrt{\hbar
/M\omega _{\perp}}$ has the dimension of a length. 
\begin{eqnarray}
Z_{n_z}(\xi) & = N_{n_z} e^{-\xi^2} H_{n_z}(\xi) \nonumber \\ & N_{n_z}=
\frac{1}{\left(\alpha_z\sqrt{\pi} 2^{n_z} n_z ! \right)^{1/2}}
\end{eqnarray}
\begin{figure*}[ht]
\includegraphics[width=5.6cm]{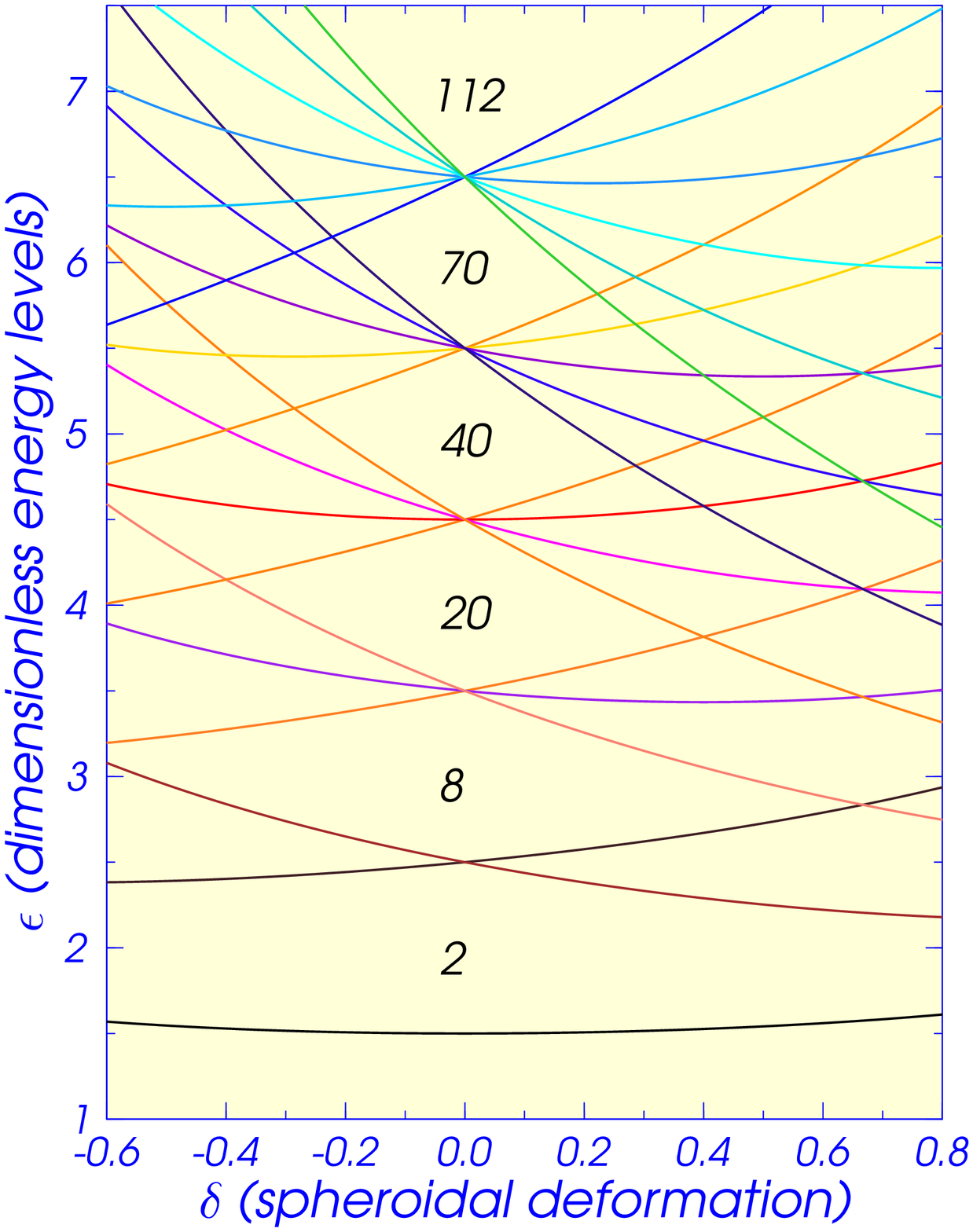} %clemen_lev} 
\hspace*{-1mm}
\includegraphics[width=5.6cm]{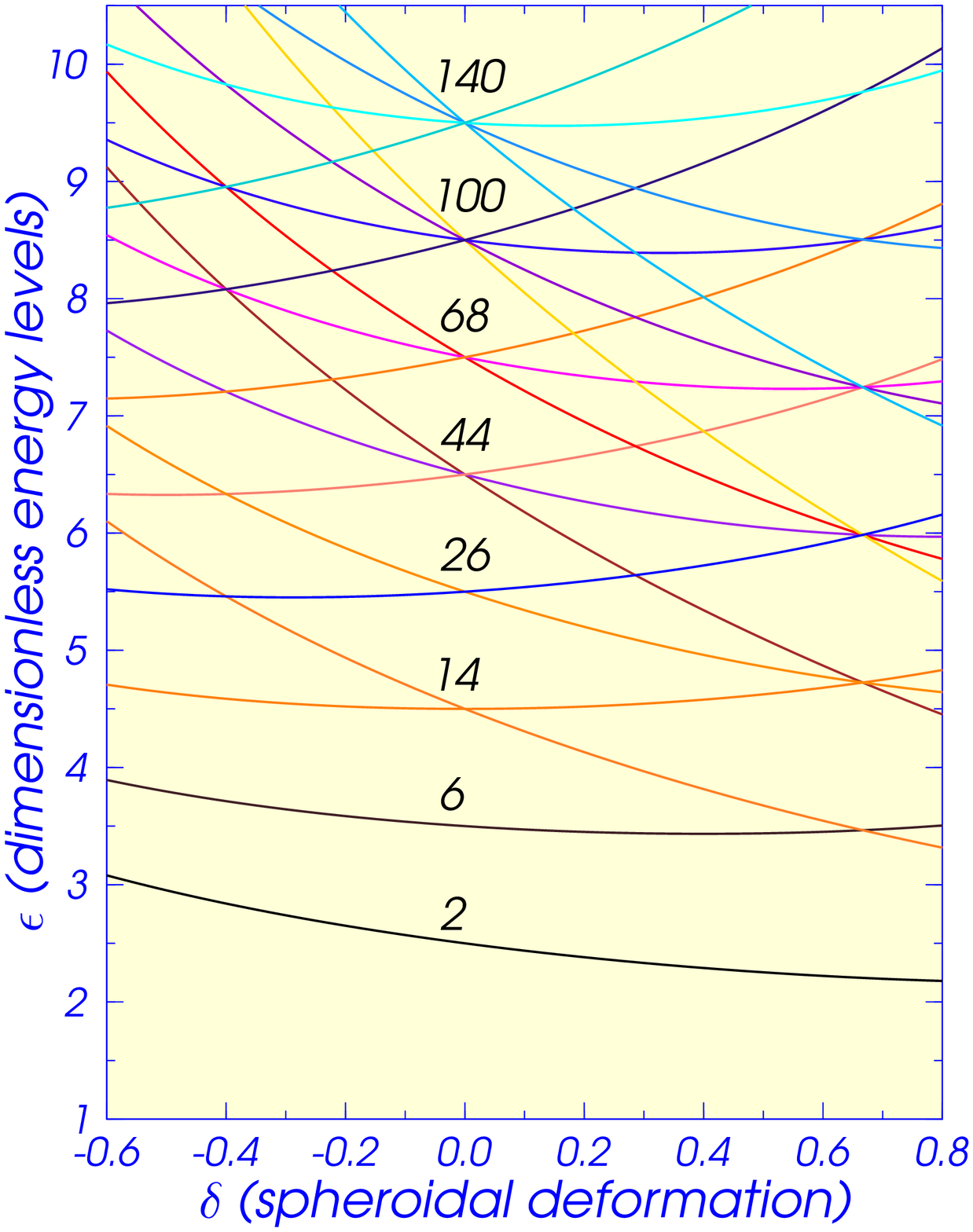} %semsph_lev}
\caption{LEFT: 
Spheroidal harmonic oscillator energy levels in units of $\hbar
\omega_0$ vs. the deformation parameter $\delta$.
Only 6 major shells ($N=0, 1, 2, ..., 5$) have been considered. Each level
is labeled by $n,n_\perp$ quantum numbers
and is $(2n_\perp +2)$-fold degenerate. The labels
are $0,0$; $1,0$, $1,1$; $2,0$, $2,1$, $2,2$; $3,0$, $3,1$, $3,2$, $3,3$; 
$4,0$, $4,1$, $4,2$, $4,3$, $4,4$, etc.
RIGHT: Semi-spheroidal harmonic oscillator energy levels in units of $\hbar
\omega_0$ vs. the deformation coordinate $\delta$.
Only 9 major shells ($N=0, 1, 2, ..., 8$) have been considered. Each level
is labeled by $n,n_\perp$ quantum numbers (with $n_z=n-n_\perp = 1, 3, 5, ...$
and is $(2n_\perp +2)$-fold degenerate. The labels
are $1,0$; $2,1$; $3,2$, $3,0$; $4,3$, $4,1$; $5,4$, $5,2$, $5,0$; 
$6,5$, $6,3$, $6,1$, etc. The semi-spherical magic numbers are identical with
those obtained at the oblate spheroidal superdeformed shape ($\delta=-2/3$):
2, 6, 14, 26, 44, 68, 100, 140, ...
\label{lev}}
\end{figure*}
where $\xi=R_0 z/\alpha_z$, $\alpha_z =\sqrt{\hbar /M\omega _z}$, and 
the main quantum number $n=n_\perp + n_z= 0, 1, 2, ...$.

The eigenvalues are
\begin{equation}
E_n=\hbar \omega_\perp (n_\perp + 1) + \hbar \omega_z(n_z+1/2)
\end{equation}

The parity of the Hermite polynomials $H_{n_z} (\xi)$ is given by $(-1)^{n_z}$
meaning that the even order Hermite polynomials are even functions $H_{2n_z}
(-\xi)=H_{2n_z}(\xi)$ and the odd order Hermite polynomials are odd
functions $H_{2n_z+1}(-\xi)=- H_{2n_z+1}(\xi)$.
There is a recurrence relationship $2zH_n=H_{n+1}+2nH_{n-1}$.
One has $H_0=1$, $H_1=2z$, $H_2=4z^2-2$, $H_3=8z^3-12z$, $H_4=16z^4-48z^2 +
12$, $H_5=32z^5-160z^3 +120z$, etc.

In units of $\hbar \omega_0$ the eigenvalues, $\epsilon=E/(\hbar \omega_0)$, 
are given by
\begin{equation} 
\epsilon  
=\frac{2}{(2-\delta)^{1/3}(2+\delta)^{2/3}}
\left[n+\frac{3}{2}+\delta \left(n_\perp -\frac{n}{2} +\frac{1}{4}\right) 
\right]
\label{sl}
\end{equation}
For a prolate spheroid, $\delta>0$, at $n_\perp=0$ the energy level decreases
with deformation
except for $n=0$, but when $n_\perp=n$ it increases.
\begin{figure}[htb]
\centerline{\includegraphics[width=8cm]{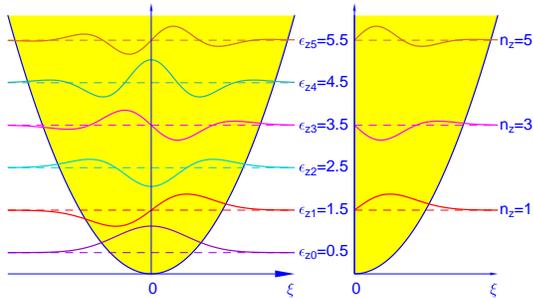}} %pot_wavef}}
\caption{LEFT: Harmonic oscillator potential $V=V(\xi)$,
the wave functions $Z_{n_z}=Z_{n_z}(\xi)$ for $n_z=0, 1, 2, 3, 4, 5$ and
the corresponding contributions to the total energy levels
$\epsilon_{z \ n_z}=E_{n_z}/\hbar\omega_z=(n_z+1/2)$ for spherical shapes, 
$\delta=0$. $\xi=zR_0/\sqrt{\hbar/M\omega_z}$.
RIGHT: The similar functions for a semi-spherical harmonic oscillator
potential. Only negative parity states are retained which are vanishing at
$\xi=0$ where the potential wall is infinitely high .}
\label{potwa}
\end{figure}
For a given prolate deformation and a maximum energy
$\epsilon_m$, there are $n_{min}$ closed shells and other levels for
\begin{figure*}[htb]
\centerline{\includegraphics[width=10cm]{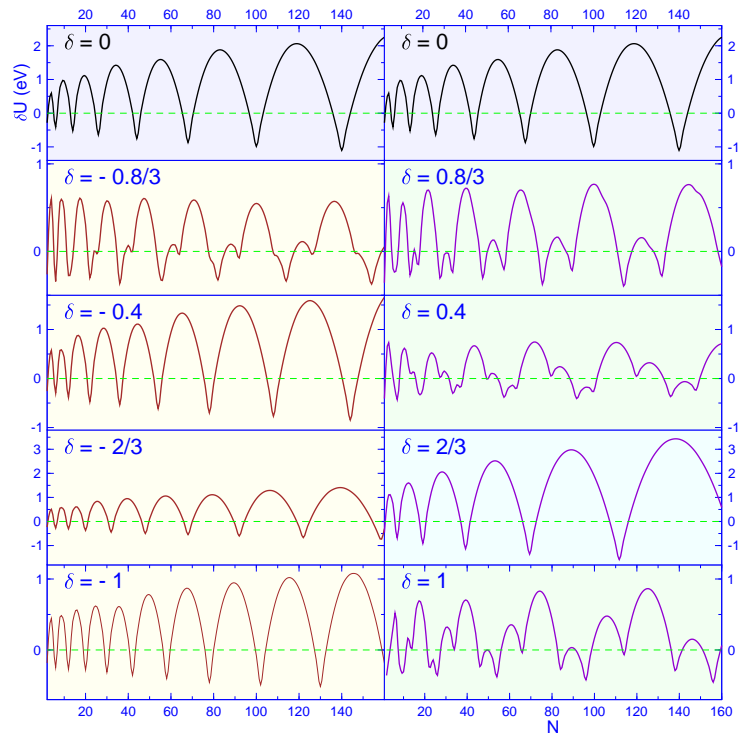}} %sh+p_sems_n5na}}
\caption{Variation of shell corrections
with $N$ for Na clusters. TOP: $\delta=0$. The semi-spherical magic numbers 
are identical with those obtained at the oblate spheroidal 
superdeformed shape: 2, 6, 14, 26, 44, 68, 100, 140, ... 
{\em For a prolate superdeformed ($\delta =  2/3$) shape the
magic numbers are identical with those obtained at the spherical shape:
2, 8, 20, 40, 70, 112, 168, ...} Other magic numbers are given in 
Table~\ref{tab}. Oblate and prolate 
shapes are considered on the left-hand side and right-hand side,
respectively.
\label{shan}}
\end{figure*}
high-order shells up to $n_{max}$:
\begin{equation}
n_{min}=\left[(2-\delta)^{1/3}(2+\delta)^{2/3}\epsilon_m-\frac{\delta}{2}
-3\right]\frac{1}{2+\delta}
\end{equation}
\begin{equation}
n_{max}=\left[(2-\delta)^{1/3}(2+\delta)^{2/3}\epsilon_m-\frac{\delta}{2}
-3\right]\frac{1}{2-\delta}
\end{equation}
and similar formulae for oblate deformations, $\delta<0$.
The low lying energy
levels for the six shells (main quantum number $n=0, 1, 2, 3, 4, 5$) can be
seen in figure~\ref{lev}. 
Each level, labelled by $n_\perp,n$, may accomodate $2n_\perp +2$ particles.
One has $2\sum_{n_\perp = 0}^n (n_\perp +1) = (n+1)(n+2)$ nucleons
in a completely filled shell charcterized by
$n$, and the total number of states of the low-lying $n+1$ shells is
$\sum_{n=0}^n (n+1)(n+2)=(n+1)(n+2)(n+3)/3$ leading to the magic numbers
$2, 8, 20, 40, 70, 112, 168 ...$ for a spherical shape.
Besides the important degeneracy at a spherical
shape ($\delta=0$), one also have degeneracies at some superdeformed shapes,
e.g. for prolate shapes at the ratio $c/a=(2+\delta)/(2-\delta)=2$ i.e.
$\delta=2/3$. More details may be found in the Table~\ref{tab}.
The first five shells
can reproduce the experimental magic numbers mentioned above; in order to describe
the other shells Clemenger introduced the term proportional to $({\bf l}^2 - 
\langle {\bf l}^2 \rangle_n)$.

Let us consider a particular shape (half of an oblate or prolate spheroid)
of a semi-spheroidal cluster deposited on a surface with the $z$ axis
perpendicular on the surface and the $\rho$ axis in the surface plane. 
Then the semi-spheroidal surface equation is given by
\begin{equation} \rho^2=\left \{
\begin{array}{ll} (a/c)^2(c^2 - z^2) & \
\ \ \ z \geq 0 \\ 0 & \ \ \ \ z < 0 \end{array} \right .
\end{equation}
The radius of the semi-sphere
obtained for the deformation $\delta=0$ is $R_s$, given by the volume
conservation, $(1/2)(4\pi R_s^3/3)=4\pi R_0^3/3$, leading to
$R_s=2^{1/3}R_0$. We shall give $\rho, z, a, c$ in units of $R_s$
instead of $R_0$. According to the volume conservation,
$a^2cR_s^3/2=R_0^3$ so that $a^2c=1$. Other kind of shapes obtained from a
spheroid by removing less or more than its half (as in the liquid drop
calculations \cite{sem07}) will be considered in the future; in this case 
it is not possible to obtain analytical solutions.

The new potential well we have to
consider in order to solve the quantum mechanical problem is shown in 
the right-hand side of the figure~\ref{potwa}.
The potential along the symmetry axis, $V_z(z)$, has a wall of an infinitely 
large height at $z=0$, and concerns only positive values of $z$
\begin{equation}
V_z=\left \{ \begin{array}{ll} \infty & \ \ \ \ z=0\\
    MR_s^2\omega_z^2z^2/2 & \ \ \ \ z \geq 0 \end{array}       \right .
\end{equation}
In this case the wave functions should vanish in the origin, where the
potential wall is infinitely high, so that only
negative parity Hermite polynomials ($n_z$ odd) should be taken into
consideration. 

From the energy levels given in figure~\ref{lev} we have to
select only those corresponding to this condition.
In this way the former lowest level with $n=0, n_\perp=0$ should be excluded. From
the two leveles with $n=1$ we can retain the level with $n_\perp=0$ i.e.
$n_z=1$. This will be the lowest level for the semi-spherical harmonic
oscillator and will accomodate $2n_\perp + 2= 2$ atoms. From the three
levels with $n=2$ only the one with $n_z=n_\perp=1$ 
\begin{table*}[bth]
\caption{TOP: Deformed magic numbers of the spheroidal harmonic oscillator.\\
BOTTOM: Deformed magic numbers of the semi-spheroidal harmonic oscillator.
}

\begin{tabular}{|p{1.5cm}|p{1cm}|p{4.5cm}||p{1.5cm}|p{1cm}|p{4.5cm}|} \hline
\multicolumn{3}{|c||}{\bf OBLATE } & \multicolumn{3}{c|}{\bf PROLATE } \\
\hline
 $\delta$ & $a/c$ & Magic numbers & $\delta$ & $a/c$ & Magic numbers \\
\hline

$-0.8/3$ & 17/13  & 2, 8, 18, 20, 34, 38, 58, 64, 92, 100, 136, 148,
... & 0.8/3 & 13/17  &2, 8, 20, 22, 42, 46, 76, 82, 124, 134 ... \\ \hline

$-0.4$ & 1.5 & 2, 6, 8, 14, 18, 28, 34, 48, 58, 76, 90, 114, 132, ... 
& 0.4 & 2/3 &2, 8, 10, 22, 26, 46, 54, 66, 84, 96, 114, 138, 156, ...  \\ \hline

$-2/3$ & 2 & 2, 6, 14, 26, 44, 68, 100, 140, ... & 2/3 & 0.5 & 
2, 4, 10, 16, 28, 40, 60, 80, 110, 140, ... \\  \hline

$-1$ & 3 & 2, 6, 12, 22, 36, 54, 78, 108, 144,  ... & 1 & 1/3 & 
4, 12, 18, 24, 36, 48, 60, 80, 100, 120, 150, ... \\  \hline

\end{tabular}

\begin{tabular}{|p{1.5cm}|p{1cm}|p{4.5cm}||p{1.5cm}|p{1cm}|p{4.5cm}|} \hline

$-0.8/3$ & 17/13  &  2, 6, 12, 22, 26, 36, 42, 56, 64, 82, 92, 114, 126, 154,
...  & 0.8/3 & 13/17  & 2, 6, 8, 14, 18, 28, 34, 48, 58, 76, 90, 114, 132,
...    \\ \hline

$-0.4$ &1.5 &2, 6, 12, 22, 36, 54, 78, 108, 144, ...&
0.4    &2/3 &2, 8, 18, 20, 34, 38, 50, 58, 64, 80, 92, 100, ... \\ \hline

$-2/3$ &2   &2, 6, 12, 20, 32, 48, 68, 92, 122, 158, ... &
2/3    &0.5 &2, 8, 20, 40, 70, 112, 168, ... \\ \hline

$-1$   &3   &2, 6, 20, 30, 42, 58, 78, 102, 130, ... &
1      &1/3 &2, 8, 10, 14, 22, 26, 46, 54, 66, 84, 96, 114, 138, 156, ...
 \\ \hline
\end{tabular}
\label{tab}
\end{table*}
\normalsize
with $2n_\perp + 2= 4$
degeneracy is retained so that the first two magic numbers at spherical
shape ($\delta=0$) are now 2 followed by 6, etc. Some deformed magic numbers
may be found in the Table~\ref{tab} and as position of
minima in Fig.~\ref{shan}. 

Each level, labelled by $n_\perp,n$, may accomodate $2n_\perp +2$ particles.
When $n$ is an odd number, one should only have even $n_\perp$ in order to
select the odd $n_z=n-n_\perp$. The contribution of the shells with odd $n$
to the semi-spherical 
magic numbers will be
\begin{equation}
\sum_{n_\perp ^{even}= 0}^{n-1}(2n_\perp + 2) = \frac{(n+1)^2}{2}
\end{equation}
leading to the sequence 2, 8, 18 for $n=1, 3, 5$. The contribution of the
shells with even $n$ to the semi-spherical magic numbers will be
\begin{equation}
\sum_{n_\perp ^{odd} = 1}^{n-1}(2n_\perp + 2) = \frac{n(n+2)}{2}
\end{equation}
which gives the sequence 4, 12, 24 for $n=2, 4, 6$. This should be
interlaced with the preceding one so that the magic numbers will be 2,
$2+4=6$, $6+8=14$, $14+12=26$, $26+18=44$, $44+24=68$, as shown at the 
right-hand side of the Fig.~\ref{lev}.

The equation (\ref{sl}) 
from the harmonic oscillator, in units of $\hbar \omega_0$
is still valid, but one should only allow the values of $n$ and $n_\perp$ for
which $n_z=n-n_\perp \geq 1$ are odd numbers.

The ortonormalization condition of the $Z_{n_z}$ component of the wave function 
became
\begin{equation} 
\int_{0}^{+\infty} Z _{n^\prime_ z}(z) Z_{n_z}(z) dz = 
\delta_{n^\prime _z n_z}  
\end{equation}
with $n_z = 1, 3, 5, ..., n$ for odd $n$ and $n_z = 1, 3, 5, ..., n-1$ for
even $n$. Consequently the normalization factor is $\sqrt{2}$ 
times the preceding one
\begin{eqnarray}
Z_{n_z}(\xi) & = \sqrt{2}N_{n_z} e^{-\xi^2} H_{n_z}(\xi) \nonumber \\ &
N_{n_z}=
\frac{1}{\left(\alpha_z\sqrt{\pi} 2^{n_z} n_z ! \right)^{1/2}} 
\end{eqnarray}

For a nucleus with mass number $A$ the shell gap is given by $\hbar \omega_0^0=
41A^{1/3}$~MeV. For an atomic cluster \cite{cle85phd} the single-particle 
shell gap is given by
\begin{equation} 
\hbar \omega_0(N) = %\frac{49 \; {\rm eV \; bohr^2}}{r^2_sN^{1/3}}\left [
\frac{13.72 \; {\rm eV \; {\mbox \AA}^2}}{r_sR_0}\left [
1+\frac{t}{r_sN^{1/3}} \right ]^{-2}
\label{shg}
\end{equation} 
which is $3.0613N^{-1/3}$ eV in case of Na clusters. 
Since we consider solely monovalent elements, $N$ in this eq. is the
number of atoms and $t$ denotes the 
electronic spillout for the neutral cluster according to \cite{cle85phd}.

The shell correction energy, $\delta U$ \cite{str67np}, in figure~\ref{shan} 
shows minima at the oblate and prolate
magic numbers given in the lower part of the table~\ref{tab}. The striking 
result is that the superdeformed prolate magic numbers
of the semi-spheroidal shape are identical with those obtained at the
spherical shape of the spheroidal harmonic oscillator.
We expect that this kind of symmetry will not be present anylonger 
for the Hamiltonian including the term proportional to 
$({\bf l}^2 - \langle {\bf l}^2 \rangle_n)$ and/or the more complex
equipotential surface we shall study in the future.

%\bibliography{clust_atom}
%\bibliographystyle{apsrev} %prsty}

\end{document}